\documentclass[aps, prd, twocolumn, showpacs, superscriptaddress, groupedaddress]{revtex4} 
\usepackage{graphicx}	
\usepackage{amssymb}
\usepackage{dcolumn}
\usepackage{color}
\usepackage{subfigure, rotating, bm, array}
\usepackage[pagebackref=false, colorlinks=true]{hyperref}
\hypersetup{
linkcolor=blue,     
citecolor=blue,     
urlcolor=blue}      
\begin{document}
\title{Thermodynamics of Reissner–Nordstr\"om black-bounce \\ black hole}
\author{Feba C Joy}
\email{febacjoy31@gmail.com}
\affiliation{Mar Thoma College, Tiruvalla, Pathanamthitta, Kerala, India}
\author{Tharanath R}
\email{tharanath.r@gmail.com}
\affiliation{Aquinas College, Edacochin, Kerala, India}

\date{\today}

\begin{abstract}
Black bounce black holes are modification of classical black hole solutions that regularize the singularity by using a bouncing parameter. In our work, we explored the thermodynamics of Reissner-Nordst\"orm black-bounce-black hole and determined thermodynamic properties such as entropy, mass, temperature, free energy and heat capacity. Along with this, we investigated the interconnection between entropy and other thermodynamic properties through graphical analysis. This provides insight into studying the extended phase space of the particular black hole. We further determined the logarithmic correction to entropy term.
\\
\\
$\boldsymbol{key words}$ : Black bounce, S V spacetime, logarithmic correction, Conformal Field Theory, P V isotherm.
\end{abstract}
\maketitle
\section{Introduction}
\label{intro}
Black holes are the regions of spacetime where nothing not even light, can escape, due to strong gravitational pull. In 1930s, Oppenheimer, Snyder and Dutt using Einstein's general relativity studied the collapse of massive stars and came to a conclusion that during the collapse of a star an event horizon is formed which is a one-way membrane through which nothing can escape. Later on, John Wheeler named this region Black hole which is the outcome of gravitational collapse described by the Oppenheimer-Snyder-Dutt(OSD) model[1,2]. \\Based on the  physical characteristics, black holes are classified into two : classical and regular. Classical black holes are the solution to Einstein field equation whereas the regular black holes arise from the theoretical models in general relativity.\par
\vspace{0.3cm}

The concept of black-bounce spacetime (Simpson Visser spacetime) was introduced by Simpson and Visser in 2019, corresponds to the geometry of a regularized black hole which is commonly known as black-bounce black hole[3]. Such black holes features a minimal surface instead of having a singularity at the center, this results in a regular black hole without infinite densities. Depending upon certain parameters, SV spacetime could describe more exotic objects like traversable wormholes. On comparing with other regular black holes these show similarities in gravitational effects as well as shadows, thus this can be regarded as mimickers to classical black holes[4]. Recent studies[5] showed, the physical origin of black bounce black holes are linked with the absorption of particles by the black hole itself. This offers a new opportunity in understanding black bounce spacetime at a deeper level[6-11]. Alongside, a charged version was proposed, known as the Reissner-Nordstr\o"m black-bounce black hole, it incorporates standard Maxwell electromagnetism with an anisotropic fluid[12].\par 
\vspace{0.3cm}
The black-bounce solutions provide valuable insights to alternative black hole models particularly in the realm of quantum gravity. This solution is mathematically regular but still replicates the key observable features of classical black holes. This makes them important candidates for further astrophysical studies. In these models, to address singularity issues the coordinate r is transformed to $\sqrt{r^2 + l^2}$, where l is the length parameter. For regular black holes l is less than 2, for traversable wormholes it's greater than 2 and for black holes with null throat l is equal to 2[13]. Similar to this regularization property, numerous black-bounce solutions exist which prefer SV spacetime to analyze their properties[14-17]. Very recently, the authors of Ref[18] reviewed the SV spacetime surrounded by the string cloud which was previously introduced by Rodrigues et al.[19], who proposed a black hole mimicker string cloud whose presence makes the black-bounce spacetime regularized. \par
\vspace{0.3cm}
{\sloppy
In addition to theoretical concepts, observational studies has further supported black hole physics. In [20] the authors concluded that the gravitational waves from the collision of two black holes were first detected by LIGO in 2015, confirmed the existence of black holes and Einstein's theory of general relativity. Subsequent observations captured a black hole neutron star merger revealing gravitational wave signals and tidal deformations [21-23]. In parallel, the Event Horizon Telescope provided the first image of a black hole shadow which arising from gravitational lensing[24]. Building on these observations, the phenomenon of gravitational lensing in the strong deflection limit has been studied in the context of black-bounce black holes[25,26].
} 
  \par
\vspace{0.3cm}
An important aspect in  black hole physics is the black hole thermodynamics. This combines general relativity with quantum mechanics and classical thermodynamics. Many studies have highlighted that black holes exhibit thermodynamic behaviors. Using general black hole metrics one can derive the thermodynamic properties[27]. With this in mind we aim to explore the thermodynamics of Reissner-Nordst\"orm black-bounce black hole.\par
\vspace{0.3cm}
This paper includes, Section(2) Thermodynamic Properties, Section (3) Logarithmic Correction in Area Law , Section (4) Equation of State and Extended Phase Transition, Section (5) Result and Discussion

\section{Thermodynamic Properties}\label{sec1}
The Reissner–Nordström black-bounce-black hole  is the solution to the electro vac Einstein field equation. It's metric is of the form[13]:\par
\vspace{0.3cm}
\begin{equation}
{ ds^2 = -f(r)dt^2+\frac{dr^2}{f(r)}+h(r)d\Omega^2_2 } 
\end{equation}
where,
\[
{f(r)= 1-\frac{2M}{\sqrt{r^2+l^2}}+\frac{Q^2}{r^2+l^2}}
\]

  and  
\[
 {h(r)=r^2+l^2}
\]
  \vspace{0.3cm}
where, M is the mass, Q is the charge and r is the event horizon radius. \par
\vspace{0.5cm} 
\begin{figure}[!htbp]

             \centering
 
    \includegraphics[width=1\linewidth]{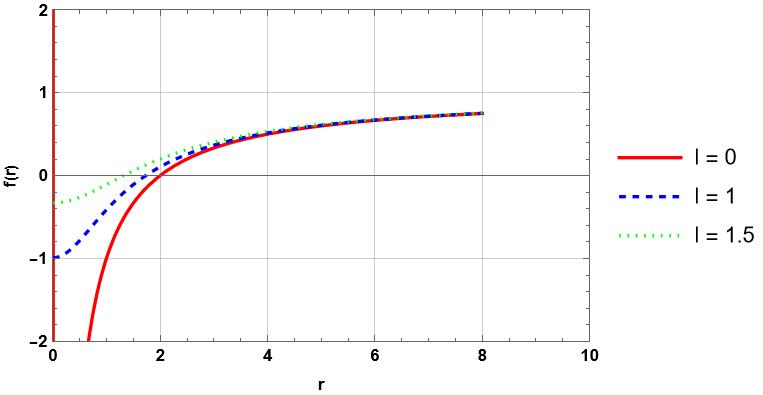}

      \caption{\textit{The figure shows the variation of \( f(r) \) with the horizon radius \( r \), where \( Q = 0.1 \). The red line illustrates the variation when \( l = 0 \), the blue line illustrates the variation when \( l = 1 \), and the green line illustrates the variation when \( l = 1.5 \).}}
     \label{fig:enter-label}
 \end{figure}
 Using equation (1) the mass \textit{M} of the given  black hole can be obtained as, 
\begin{equation}
    {M} = \frac{Q^2+r^2+l^2}{2\sqrt{r^2+l^2}}
\end{equation}
   \vspace{0.3cm}
From the first law of black hole thermodynamics we have $$ dM = T_H dS $$ This results in the Bekeinstein - Hawking  Entropy,
\[
S = \frac{A}{4 l_p ^2}
\]
           
where A is the surface area of the black hole horizon and $l_p$ is the Planck length.  \par
\vspace{0.3cm}
Therefore the entropy  S can be expressed as a function of  r as,
\[
r=\sqrt{\frac{S}{\pi}}
\]
\par
\vspace{0.3cm}
Applying the above relations, the mass function may be expressed as, 
\begin{equation}
           \textbf{M} = \frac{Q^2+\frac{S}{\pi}+l^2}{2\sqrt{\frac{S}{\pi}+l^2}}
       \end{equation}
       \begin{figure}[!htbp]
\centering

    \includegraphics[width=1\linewidth]{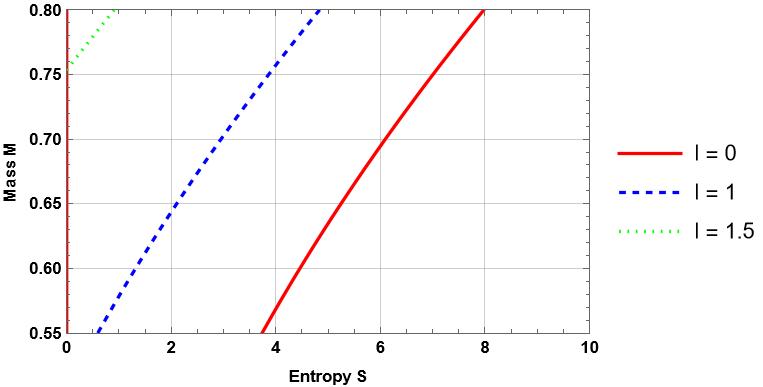}

           \caption{\textit{The figure shows the variation of Mass \( M \) with Entropy \( S\), where \( Q = 0.1 \). The red line illustrates the variation when \( l = 0 \), the blue line illustrates the variation when \( l = 1 \), and the green line illustrates the variation when \( l = 1.5 \).}}
    \label{fig:enter-label}
\end{figure}
 \par
\vspace{0.3cm}
The other thermodynamic quantities such as temperature, heat capacity and free energy are deduced from the standard thermodynamic relations.
        
    \vspace{0.5cm}
    Temperature of the black hole is given by the relation,
   \[
     T = \frac{\partial M}{\partial S}
   \]
Thus, we obtain the temperature as :
\begin{equation}
   \textbf{ T} =\frac{2(\frac{S}{\pi}+l^2) - Q^2 +\frac{S}{\pi}+l^2}{4\pi (\frac{S}{\pi}+l^2)^{3/2}}
\end{equation}
\vspace{-0.5cm}
\begin{figure}[!htbp]
\centering

           \includegraphics[width=1\linewidth]{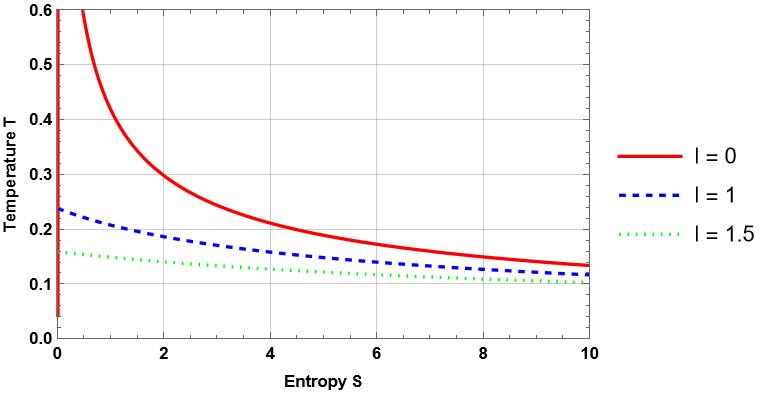}
      
              \caption{\textit{The figure shows the variation of Temperature \( T\) with Entropy \( S\), where \( Q = 0.1 \). The red line illustrates the variation when \( l = 0 \), the blue line illustrates the variation when \( l = 1 \), and the green line illustrates the variation when \( l = 1.5 \).}}
    \label{fig:enter-label}
\end{figure}
\par
\vspace{0.5cm}
Fig(3) indicates a smooth variation of temperature with entropy, which excludes first order phase transition. Hence we calculate the heat capacity as,
\[
 C = T\frac{\partial S}{\partial T}
\]
   From the general relation heat capacity can be\\ expressed as,
\begin{equation}
    \textbf{C} = \frac{2(S+l^2\pi-Q^2 \pi)}{(3Q^2+\frac{3S}{\pi}+3l^2)(\frac{S}{\pi}+l^2)-4(\frac{S}{\pi}+l^2)^3}
\end{equation}
 \vspace{0.3cm}
  \begin{figure}[!htbp]
    \centering
   
     \includegraphics[width=1\linewidth]{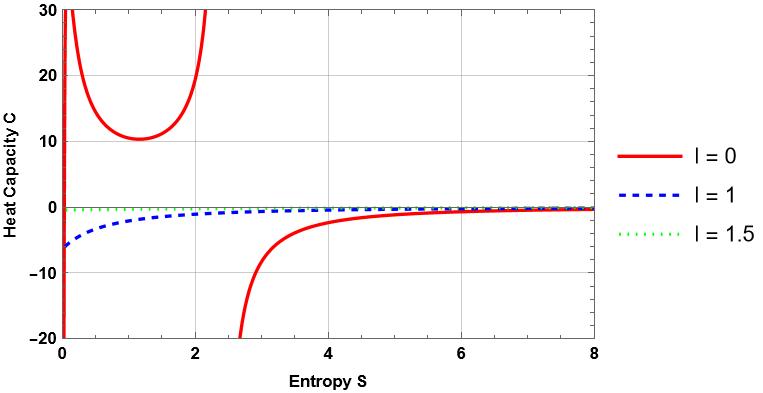}

    \caption{\textit{The figure shows the variation of Heat Capacity \( C\) with the Entropy \( S \), where \( Q = 0.1 \). The red line illustrates the variation when \( l = 0 \), the blue line illustrates the variation when \( l = 1 \), and the green line illustrates the variation when \( l = 1.5 \).}}
    \end{figure}
    \par
    \vspace{0.5cm}
     Using the above thermodynamic quantities, we can now analyze the free energies of the particular black hole. In this work, the Hawking temperature and thermodynamic temperature are regarded as distinct quantities. The Helmholtz free energy is calculated using thermodynamic temperature T indicating the local thermodynamic effects, while the Gibbs free energy is evaluated using Hawking temperature, representing the global thermodynamic behavior.
   The equations for Helmholtz Free Energy and Gibbs Free energy for the Reissner-Nordst\"orm black-bounce black hole  are given as follows : \par
   \vspace{0.3cm}
The Helmholtz Free energy is of the form,
   \[
     F = M-TS
   \]
 \begin{equation}
 \textbf{F} = \frac{Q^2\pi(S+2\pi l^2)-S^2\pi-2Sl^2\pi+2\pi^2l^4}{4\pi^2(\frac{S}{\pi}+l^2)^{3/2}}
\end{equation}
\vspace{-0.5cm}
\begin{figure}[!htbp]
 \centering

      \includegraphics[width=1\linewidth]{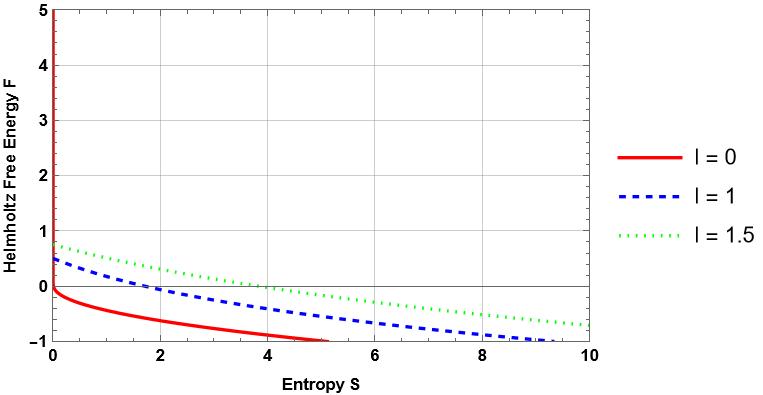}
 
    \caption{\textit{The figure shows the variation of free energy \( F\) with the Entropy \( S \), where \( Q = 0.1 \). The red line illustrates the variation when \( l = 0 \), the blue line illustrates the variation when \( l = 1 \), and the green line illustrates the variation when \( l = 1.5 \)}}
        \label{fig:enter-label}
    \end{figure}
     and the Gibbs free energy,[26],
   \[
    G = E-T_HS-\phi_{H}Q 
   \] 
     where, E is the energy which corresponds to the black hole mass, $T_H$ is the Hawking Temperature ,
 \[
  T_H =K/2\pi = \frac{r_+-r_-}{4\pi r_+^2}
  \]
  \[
   =\frac{2}{\sqrt{-l^2+2M^2-Q^2-2\sqrt{M^4-M^2Q^2 }}}
  \]
        and $\phi_H $ is the electrostatic potential on the horizon,
   \[
       \phi_H  = \frac{Q}{r}
   \]
   Now from the given relations we have the Gibbs Free energy as,
     \begin{equation}
         \textbf{G} = M - \frac{2S-Q^2}{\sqrt{l^2+2M^2-Q^2-2\sqrt{M^4-M^2Q^2 }}}
     \end{equation}
\vspace{-0.5cm}
\begin{figure}[!htbp]
  
        \centering

                       \includegraphics[width=1\linewidth]{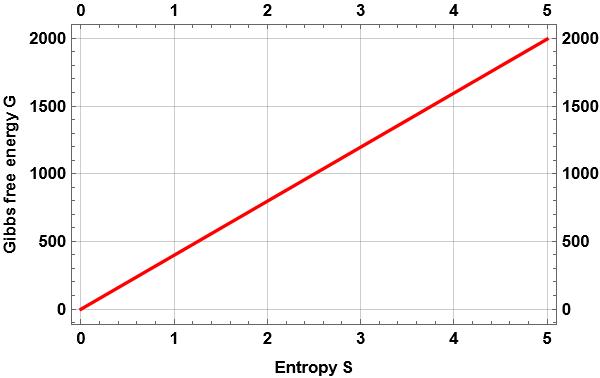}
                
  \caption{\textit{The figure shows the variation of Gibbs free energy \( G\) with the Entropy \( S \), where \( Q = 0.1 \). The red line illustrates the variation when \( l = 0 \), the blue line illustrates the variation when \( l = 1 \), and the green line illustrates the variation when \( l = 1.5 \). But all three lines overlaps indicating an identical variation of Gibbs free energy with entropy.}}
        \label{fig:enter-label}
    \end{figure}
\par
\vspace{0.5cm}

\section{Logarithmic Correction in Area Law}
\label{sec2}
The statistical fluctuations around the thermal equilibrium of black holes resulted in the logarithmic correction to entropy. These are known to be directly connected to the anomalies for a gravitational theory with quantum correction[27].
Since the logarithmic corrections to entropy arise from the contributions of Thermal Heat Capacity (THC) and Conformal Field Theory (CFT)[28] the resulting entropy relations can be expressed as follows:\par
\vspace{0.5cm}
\begin{equation}
    S_{THC} = S - \frac{1}{2}\ln|CT^2|
\end{equation}
\begin{equation}
    S_{CFT} = S-\frac{1}{2}\ln|ST^2|
\end{equation}
From equations (8) and (9)we get,
\begin{equation}
    S_{THC} = \pi r^2 -\frac{1}{2}\ln|X|
\end{equation}
where,
\[
 X = \left(\frac{2(S+l^2\pi-Q^2 \pi)}{(3Q^2+\frac{3S}{\pi}+3l^2)(\frac{S}{\pi}+l^2)-4(\frac{S}{\pi}+l^2)^3}\right)
\]
Now 
\begin{equation}
     S_{CFT}=\pi r^2-\frac{1}{2}\ln|(Y_1 + Y_2) Z|
\end{equation}
where,
\[
 Y_1 = \frac{2\pi(2M^2-Q^2-l^2)}{2}
\]

and 
\[
 Y_2 = \frac{\sqrt{4\pi M^2-2Q^2-2l^2-4\pi^2 (Q^2 - l^2)-16 M^2l^2}}{2}
\]

and 
\[
 Z =   \left(
\frac{2(\frac{S}{\pi}+l^2) - Q^2 +\frac{S}{\pi}+l^2}{4\pi (\frac{S}{\pi}+l^2)^{3/2}}\right)^2
\]

Substituting the above relations to equations (10) and (11) we get,
\begin{equation}
    \frac{S_{THC}}{S_{CFT}} = \frac{\pi r^2-\frac{1}{2}ln X}{\pi r^2 - \frac{1}{2}(ln(Y_1+Y_2)Z)}
\end{equation}
 \begin{figure}[!htbp]

    \centering

\includegraphics[width=1\linewidth]{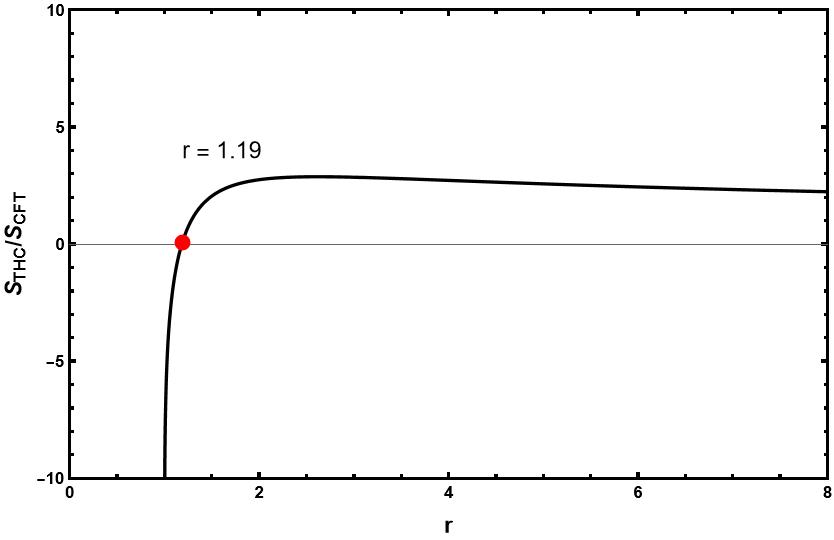}
 
   \caption{\textit{This figure shows how the ratio of logarithmic  corrections in entropy varies  with horizon radius, the dot is the point in which r =1.19}}
          \label{fig:}
      \end{figure}
\vspace{0.3cm}
Examining the fig(7), we observe that for the lower r values the ratio is less than one and it reaches unity at r = 1.19, beyond this point the ratio exceeds one. Hence for black holes with lower r values the quantum effects dominate. In other words, as the size of the black holes decreases, its event horizon becomes increasingly susceptible to quantum fluctuations. Meanwhile, for larger sizes, thermal properties play a crucial role. As the black hole's radius increases, it can store more energy and interact more extensively with its surroundings, leading to a greater thermal contribution to its entropy. Hence, we conclude that the quantum effects tend to dominate at smaller scales, whereas the thermal properties become more influential as the black hole grows in size.
\section{Equation of State and Extended Phase Transition}
\label{sec3}
The equation of state is based on the first law of black hole thermodynamics. It relates quantities such as temperature, volume and pressure similar to how the ideal gas law relates them in classical thermodynamics[29].\par
  \vspace{0.5cm}
Let the pressure P be,\par

\[
  P= -\frac{\Lambda}{8\pi}
\]

 where,$-\Lambda$=$\frac{3}{l^2} $[30]. Now,\par
\begin{equation}
    P=\frac{3}{8\pi l^2}
\end{equation}
The equation for thermodynamic volume is, 
\[
  V =\left( \frac{\partial M}{\partial P}\right)_{S,Q}
\]
Using Legendre transformation we can obtain the thermodynamic volume of a black hole in massive gravity and is given by,
  \[
   V = \frac{4}{3}\pi r^3
  \]
From the above equation,
 \[
   S = \left(\frac{3}{4}V {\pi}^{\frac{1}{2}}\right)^{\frac{2}{3}}
 \]
  Enthalpy of a black hole be,
\[
  H = E + PV
\]
and the total energy E is given by,
\[
     E = M - PV
\]
On comparing the above relations it is clear that the enthalpy is regarded the  mass of a black hole i.e,
\[
   H(S,P) = M
\]
Thus from equation (8) we get,
\begin{equation}
    H(S,P) = \frac{Q^2+\frac{S}{\pi}+\frac{3}{8\pi P}}{2\sqrt{\frac{S}{\pi}+\frac{3}{8 \pi P}}}
\end{equation}
The temperature of the Reissner–Nordström black-bounce-black hole in terms of enthalpy and entropy can be obtained by using the relation,   
\[
 T = \left(\frac{\partial H}{\partial S}\right)_P
\]
\begin{equation}
T = \frac{8PS+3-8\pi P Q^2}{32\pi^2 P \left(\frac{S}{\pi}+\frac{3}{8 \pi P}\right)^{\frac{3}{2}}}
    \end{equation}
The equation (15) indicates the equation of state of the Reissner–Nordström black-bounce-black hole.  
\vspace{0.5cm}
 \begin{figure}[!htbp]

                    \centering
                    
                    \includegraphics[width=1\linewidth]{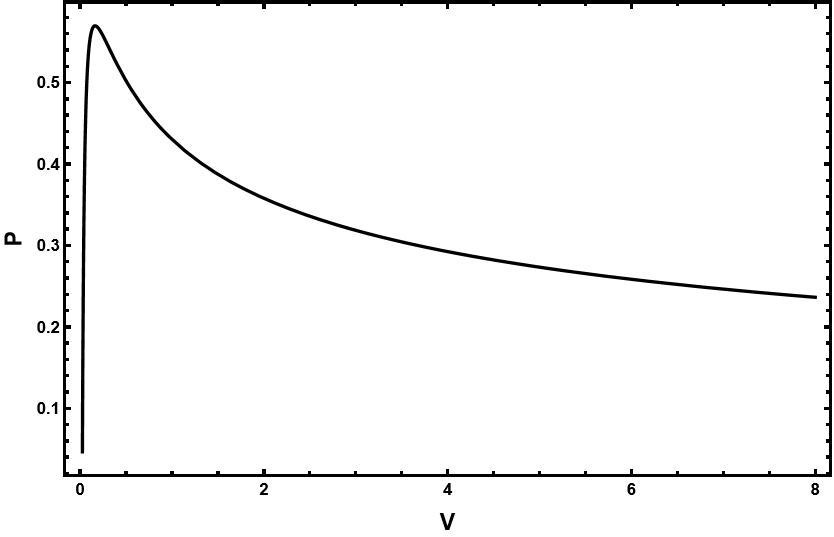}

                    \caption{\textit{The figure illustrates the PV isotherm for T=1,l=0.1,Q=0.01}}
          \label{fig:enter-label}
      \end{figure}
\par
\vspace{0.5cm}
From fig(8), it is clear that the phase transition most likely represents a single phase black hole which is similar to Hawking - page transition. Moreover, higher pressure at lower volume range suggest a quantum gravity correction.
\section{Result and Discussion}
\label{result}
In this work, We analyzed the thermodynamics of Reissner–Nordström black-bounce-black hole. At first, we obtained the thermodynamic properties including mass, temperature, heat capacity, Helmholtz free energy and Gibbs free energy and plotted them against entropy by considering Q = 0.1, and l = 0,1,2. From the temperature entropy graph the chance for first order phase transition was eliminated. As a result, heat capacity entropy graph was plotted and this clarifies that at specific entropy values, there is a clear divergence which is a key signature to second order phase transition. A transition from positive to negative or vice versa across the divergence indicates a change in stability at the transition point. Thus these rules out first order phase transition and confirms second order transition. \par
 \vspace{0.2cm}
 Moreover, upon analyzing the parametric plots we realize that, the variation of mass with temperature indicates a continuous profile without any abrupt change. The absence of inflection points supports the  non existence of first order phase transition. Similarly, the mass free energy graph reveals a smooth and monotonic relation, with both evolving continuously. This behavior confirms the absence of first order transition. Finally, the graph showing the variation of heat capacity with horizon radius indicates a divergence and a change in sign at a critical radius. Such a divergence confirms a second order transition in black hole thermodynamics and the change in sign separates stable and unstable thermodynamic phases of the particular black hole.
    \begin{figure}[!htbp]

              \centering

                  \includegraphics[width=1\linewidth]{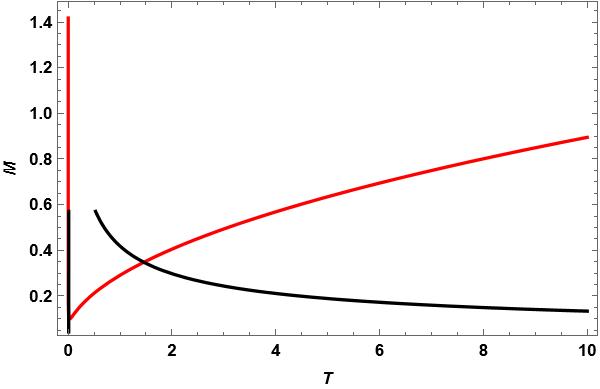}

                \caption{\textit{Figure shows the extrapolated graph of Mass (black line) and Temperature (red line)here Q=0.1 and l=0}}
        \label{fig:enter-label}
    \end{figure}

\par
\vspace{0.5cm}
 \begin{figure}[!htbp]
    \centering

                                 \includegraphics[width=1\linewidth]{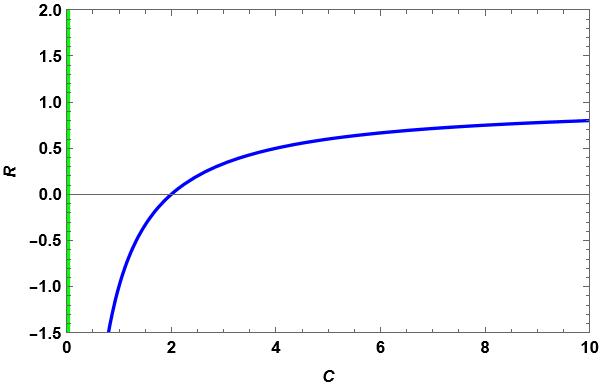}

 \caption{\textit{Figure shows the extrapolated graph of Heat capacity (blue line)and horizon radius (green line)here Q=0.1 and l=0}}
        \label{fig:enter-label}
    \end{figure}
     \begin{figure}[!htbp]
    \centering

         \includegraphics[width=1\linewidth]{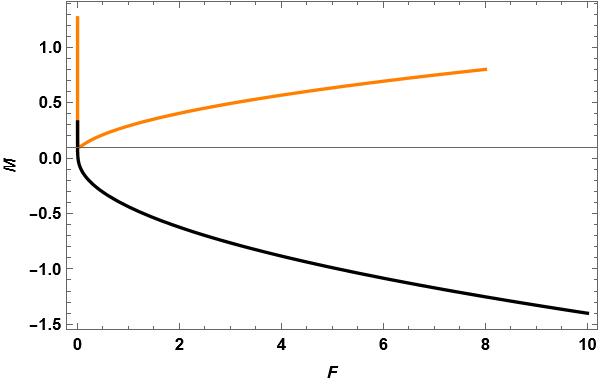}
   
          \caption{\textit{Figure shows the extrapolated graph of Mass (orange line) and Free energy (black line),Q=0.1 and l=0}}
        \label{fig:enter-label}
    \end{figure}
 \par
 \vspace{0.5cm}
 \begin{flushleft}
     On further, we introduced an entropy correction in thermal heat capacity and  conformal field theory. This reveals that the quantum effect dominates at smaller scales while, thermal properties become more influential for larger scales. We subsequently studied the P-V isotherm. Upon examination, it is evident that the graph exhibits a typical inverse relationship between pressure and volume at constant temperature indicating an isothermal process. In conclusion, the absence of  oscillations or inflection points suggests that there may not be a first-order phase transition.
 \end{flushleft}

\end{document}